\documentclass[12pt]{article}
\usepackage{fullpage,epsfig,graphics,amsbsy,amssymb,cancel,slashed,mathrsfs}
\usepackage{psfrag,hyperref}
\usepackage{graphicx,color}
\usepackage{wrapfig}

\newcommand{\be}{\begin{eqnarray}}
\newcommand{\ee}{\end{eqnarray}}
\usepackage{amsmath}
\numberwithin{equation}{section}

\newcommand{\bea}{\begin{eqnarray}}
\newcommand{\eea}{\end{eqnarray}}  
\newcommand{\nn}{\nonumber}
\newcommand{\Tr}{\textrm{Tr}}

\newcommand{\NN}{\mathcal{N}}
\newcommand{\la}{\lambda}
 \newcommand{\sfrac}[2]{\mbox{$\frac{#1}{#2}$}}
\newcommand{\ms}{\!-\!}
\newcommand{\ps}{\!+\!}

\def\Gc{\Gamma}

\def\gs{\sigma}

\def\gl{\lambda}

\begin{document}

\thispagestyle{empty}
\begin{flushright} \small
UUITP-07/13
 \end{flushright}
\smallskip
\begin{center} \LARGE
{\bf  5D super Yang-Mills theory and the correspondence to AdS$_7$/CFT$_6$}
 \\[12mm] \normalsize
{\bf  Joseph A. Minahan, Anton Nedelin and Maxim Zabzine} \\[8mm]
 {\small\it
  Department of Physics and Astronomy,
     Uppsala university,\\
     Box 516,
     SE-75120 Uppsala,
     Sweden\\
   }
\end{center}
\vspace{7mm}
\begin{abstract}
 \noindent
We study the relation  between 5D super Yang-Mills theory and the holographic description of 
 6D  $(2,0)$  superconformal theory.  
  We start by clarifying  some issues related to the localization of $\NN=1$ SYM  with matter on $S^5$.  We  concentrate on the case of a single adjoint hypermultiplet with a mass term and  argue that the theory has a symmetry
   enlargement  at mass $M=1/(2r)$, where $r$ is the $S^5$ radius. However, in order to have a well-defined localization locus it is necessary to rotate $M$ onto the imaginary axis, breaking  the enlarged symmetry.  Based on our prescription, the imaginary mass values are physical and we show how the localized path integral is consistent with earlier results for 5D SYM in flat space.   We then compute the free   energy and the expectation value for a circular Wilson loop in the large  $N$ limit.   The Wilson loop calculation shows a mass dependent constant rescaling between weak and strong coupling.  The Wilson loop continued back to to the enlarged symmetry point is consistent with a supergravity computation for an M2 brane using the standard identification of the compactification radius and the 5D coupling.  If we continue back to the physical regime and use this value of the mass to determine the compactification radius, then  we find   agreement between the SYM free energy and the corresponding supergravity calculation. We also verify numerically some of our analytic approximations.    \end{abstract}

\eject
\normalsize

\section{Introduction and main results}

The 6-dimensional $(2,0)$  theories are the most mysterious in Nahm's classification of superconformal field theories \cite{Nahm:1977tg}.  They are maximally supersymmetric with no dimensionless parameter that can be varied between weak and strong coupling.  Their fields are comprised of self-dual tensor multiplets and  do not have a  Lagrangian description. They are conjectured to  be dual to $M$-theory on an $AdS_7 \times S^4$  background, which reduces to supergravity in the large $N$ limit.  One of the most striking results learned from the supergravity duals is an $N^3$ dependence in their  free energy and conformal anomaly \cite{Klebanov:1996un,Henningson:1998gx}.
 
 Compactification of the $(2,0)$ theory on a circle with radius $R_6$ reduces it to  5-dimensional maximally supersymmetric Yang-Mills (MSYM) \cite{Seiberg:1997ax,Witten:1995ex}, with the coupling identified with the radius by  \footnote{We normalize the Yang-Mills Lagrangian as in \cite{Pestun:2007rz}, which differs from the standard normalization by a factor of 2, and changes the relation between $R_6$ and $g_{YM}$ accordingly.}
 \be\label{R6gym}
 R_6=\frac{g_{YM}^2}{8\pi^2}\,.
 \ee
 This relation follows from the identification of the Kaluza-Klein modes in the $(2,0)$ theory with the instanton particles in the 5D MSYM \cite{Witten:1995ex}.   
  
 Recently, it has been argued that the 5D MSYM can be used to define the $(2,0)$ theories \cite{Douglas:2010iu,Lambert:2010iw,Bolognesi:2011nh}.  At first blush this seems rather strange, since  5D MSYM is not renormalizable  and hence would appear to need extra degrees of freedom for a UV completion.  In fact these extra degrees of freedom would seem to account for the $N^3$ dependence in the $(2,0)$ free   energy and anomaly.  In \cite{Douglas:2010iu} it was proposed that perhaps the 5D MSYM is actually finite, negating the need for extra degrees of freedom.  If this were true then it would be consistent for the 5D theory to have all degrees of freedom found in the $(2,0)$ theory.  However, a recent study shows that the 5D MSYM is explicitly divergent at six loops \cite{Bern:2012di} and hence requires a UV completion.
 
 Nonetheless,  in \cite{Kim:2012av,Kallen:2012zn} it was shown using localization techniques that certain 5D $\NN=1$ SYM theories on $S^5$ with 8 supersymmetries exhibit $N^3$ behavior in their free energies.  The necessary ingredient for the $N^3$ behavior is the presence of a single hypermultiplet in the adjoint representation \cite{Kallen:2012zn}, where the hypermultiplet mass sets the  overall coefficient in front of $N^3$.  We will call these $\NN=1^*$ theories.  
 
 In flat space a single massless hypermultiplet reproduces the field content of the 5D MSYM.  Adding the hypermultiplet increases the number of supersymmetries from 8 to 16 and enhances the $R$-symmetry from $SU(2)\times U(1)$ to $SO(5)$.  But in the Euclidean version of this theory  on $\mathbb{R}^5$, the $R$-symmetry must itself be rotated to the noncompact $SO(4,1)$ in order to preserve the reality conditions for the spinors.   This follows from the same dimensional reduction argument used in the study of 4D $\NN=4$ SYM \cite{Pestun:2007rz}.
 
 However, unlike 4D $\NN=4$ SYM,  5D MSYM is not conformal, hence there is no canonical way to map it from $\mathbb{R}^5$ to  $S^5$.   In \cite{Hosomichi:2012ek} it was shown how to put $\NN=1$ SYM with arbitrary hypermultiplet content on $S^5$ by adding additional terms to the Lagrangian.  On $S^5$, the noncompact $R$-symmetry of an $\NN=1$ theory is broken to $SO(1,1)\times U(1)$.  For the theory with a massless adjoint  hypermultiplet there is still only $\NN=1$ supersymmetry.   
  However, the authors in \cite{Kim:2012av} observed that the global symmetry is enhanced when the hypermultiplet mass is $M=1/(2r)$, where $r$ is the $S^5$ radius.  They also provided evidence that the supersymmetry is increased to 16 supersymmetries, suggesting that the enhanced global symmetry is an $R$-symmetry.  A further argument is that   by inserting this mass into the expression for the localized path integral there is    a vast simplification \cite{Kim:2012av}, reminiscent of  4D $\NN=2^*$ SYM on $S^4$, where the determinants in the path integral take a very simple form and  all instanton factors cancel out  at the $\NN=4$ point\footnote{We thank V. Pestun for a discussion on this point.}.
 
 Ultimately, we want to compare the free energy from 5D SYM with an analogous computation for the $(2,0)$ theory using its  supergravity dual.   It is here where the situation is problematic, namely because there does not exist a Euclidean version of $(2,0)$ theory.  The argument for this is simple.  In the Lorentzian case one can have 16 real spinors by combining an  $SO(1,5)$ Weyl spinor with  the spinor of the  $SO(5)$ $R$-symmetry. However, if we attempt a Euclidean rotation then charge conjugation maps an $SO(6)$ spinor to the other spinor representation, meaning that we have to split the $R$-symmetry spinors.  But this is only possible if we split the $SO(5)$ $R$-symmetry into a noncompact version of $SU(2)\times SU(2)$.  What we are left with is the Euclidean version of the 6D $(1,1)$ theory, which can be reduced to 5D MSYM, but with a completely different relation between the compactification radius and the 5D coupling.
 
We can then  consider the $(2,0)$ theory after conformally mapping it to $\mathbb{R}\times S^5$.  It is natural to identify this $S^5$  with the $S^5$ for the 5D SYM.  However, the $\mathbb{R}$ is time-like and  requires a Euclidean rotation in order to compactify it and identify it as in (\ref{R6gym}).   Furthermore, after restricting to 16  supersymmetries  to preserve the $S^5$,  the reduced   $R$-symmetry of the $(2,0)$ theory is $SU(2)\times U(1)$, which also requires some sort of  Euclidean rotation in order to identify it with the non-compact $R$-symmetry in Euclidean 5D SYM.   So it would seem that we would run into the same sort of problem with the spinors.

Our approach to the problem is to just go ahead with the Euclidean rotations and see what we get.  What we find is that we can reach agreement between the free   energy of the 5D theory and the supergravity computation provided that we: 1) allow for a strong coupling renormalization of the coupling that we will  explain below; 2) rotate the mass parameter $M$ to $i/(2r)$.    In fact,   strictly speaking it is necessary to rotate the mass onto the imaginary axis in order to localize the path integral.

 Localization is a powerful technique that reduces the path integral to a matrix model \cite{Kallen:2012cs,Hosomichi:2012ek,Kallen:2012va,Kim:2012av}. Using localization,   the $\NN=1^*$ free   energy at  strong coupling is found to be \cite{Kallen:2012zn}
\be\label{FEI}
F=-\frac{(9/4+m^2)^2}{96 \pi }\,\la\, N^2\,,
\ee 
where  $\la=g_{YM}^2N/r$ is the  't Hooft parameter and $m\equiv i\,rM$.
Given the scaling behavior of $\la$, we see that $F$ scales as $N^3$.  This result can be compared to a supergravity calculation on $AdS_7\times S^4$ with the $AdS_7$ boundary chosen to be $S^5\times S^1$.  In this case,  using the techniques in \cite{Balasubramanian:1999re,Emparan:1999pm,deHaro:2000xn,Awad:2000aj}, the regulated total action is found to be
\be\label{IAdS}
I_{AdS}=-\frac{5\pi R_6}{12\,r}N^3~.\label{sugra-final}
\ee
Using the identification in (\ref{R6gym}), we see that there is a mismatch of $81/80$  between $F$ and $I_{AdS}$ at $m=0$, or $4/5$  if we continue (\ref{FEI})  back to the symmetry enlargement point at $m=i/2$.  We emphasize that we are comparing strong coupling results, so this mismatch is not like the case of $\NN=4$ SYM at finite temperature with its   famous factor of $3/4$ \cite{Gubser:1996de}.  
 

  While  localization can only be applied in  a  limited number of situations,  one of these is a supersymmetric Wilson loop which wraps an $S^1$-fiber of $S^5$, where the $S^5$ is  seen as the Hopf vibration  over a $CP^2$. As an example, we can consider a supersymmetric Wilson loop
that wraps the equator of the $S^5$.  At weak coupling one can show that 
\be
\langle W\rangle\sim \exp\left(\frac{\lambda}{ 8 \pi}\right)~.
\ee
In this paper, we will show that at strong coupling the Wilson loop behaves as
\be\label{W1}
\langle W\rangle\sim \exp\left((9/4+m^2)\frac{\lambda}{ 8\pi}\right)~.
\ee
  In this paper we will also compute the regularized circular Wilson loop in supergravity, which is found by wrapping an M2 brane around an $S^1$ and attaching it to a great circle on the $S^5$ boundary.   Here we find that
\be\label{W2}
\langle W\rangle_{AdS}\sim\exp\left(\frac{2\pi N R_{6}}{r}\right)~.
\ee
%
Hence, at $m=i/2$ we see that (\ref{W1}) and (\ref{W2}) are consistent with the identification in (\ref{R6gym}).

This still leaves the mismatch in the free energy. 
However, if we   rotate back to $m =1/2$  and use this value  to match $R_6$ to the coupling in (\ref{W1}) and (\ref{W2}),  we  have
 \be\label{R6gymN2_1/2}
 R_6=\frac{5g_{YM}^2}{ 32 \pi^2}\,.
 \ee
Substituting this value of $R_6$ into (\ref{sugra-final} we then  find agreement with the supergravity computation! The mass parameter can be thought of as the expectation value of a real scalar field that is part of a vector multiplet.  Localization reduces the path integral to a matrix integral over $N$ real scalars that are also part of vector multiplets.  But convergence of the integral requires a Euclidean rotation for all of these scalars.  So perhaps it is not surprising that we should also rotate the mass parameter.

The rest of the paper is organised as follows. In 
section \ref{actions-5Dtheory} we review the formulation of $\NN=1$ 5D Yang-Mills theory on $S^5$ and discuss its symmetries.  In section \ref{S-general-matrix} we present the details of the matrix model resulting from localization and explore its  different limits.  In section \ref{adjoint-5D-YM} we consider  the large $N$ behavior of the  $\NN=1^*$ theories. 
Here  we calculate the free energy  as well as the expectation value 
  of a supersymmetric  Wilson loop in the weak and strong coupling limits.  We also generalize these results to a $\mathbb{Z}_k$ quiver theory.   In section \ref{numerics} we collect  some numerical results about this model, showing that they agree with the analytical results of the previous section. In section \ref{s-sugra} we describe 
   the supergravity derivation of the free energy and Wilson loop, in particular showing that the Wilson loop is consistent with our identification of $R_6$ to $g_{YM}^2$. 
   In section \ref{summary} we discuss some open issues. 

\section{$\NN=1$ $5D$ Yang-Mills with matter on $S^5$}
\label{actions-5Dtheory}

In this section we review the construction of $\NN=1$ supersymmetric Yang-Mills theory with matter  on the five sphere $S^5$. This model has been constructed in \cite{Hosomichi:2012ek} and we follow
 their conventions. 

 Let us start with the discussion of the vector multiplet. Preserving 8 supercharges one may construct 
the  $\NN=1$  theory with the following Lagrangian density on $S^5$  
\bea&& L_{vector}= \frac{1}{g_{YM}^2} \Tr\Big[\frac{1}{2}F_{mn} F^{mn} -D_m\gs  D^m\gs-\frac12 D_{IJ}D^{IJ}+\frac{2}{r} \gs t^{IJ}D_{IJ}- \frac{10}{{r}^2}
 t^{IJ}t_{IJ}\gs^2\nn\\
&&\hspace{2cm}+i\gl_I\Gc^mD_m\gl^I-\gl_I[\gs,\gl^I]-\frac{i}{r}t^{IJ}\gl_I\gl_J\Big]~,\label{action_vector}
\eea
 where we have chosen to write the radius $r$ of $S^5$ explicitly\footnote{The unusual sign for the $\gs$ kinetic term follows the convention in \cite{Pestun:2007rz}.   We ultimately want to consider Euclidean Yang-Mills and yet work with physical fermions.  This is accomplished by making $\gs$ time-like.  This can be seen directly for the $\NN=2$ gauge theory in flat space $\mathbb{R}^5$ 
  which can be reduced from 10D Yang-Mills on $\mathbb{R}^{1,9}$.  The scalars in the 5D theory correspond to the gauge field components along the compactified directions.  Choosing one compactified direction to be time-like so that the remaining theory is Euclidean results in a wrong sign kinetic term for that scalar.}. This $\NN=1$ Yang-Mills 
  theory is not conformally invariant and  requires some guess work to construct the theory.  
 If we integrate   out the auxiliary field $D$ and consider the bosonic part of the Lagrangian
  \bea
 L_{vector} = \frac{1}{g_{YM}^2} \left [ \frac{1}{2} F_{mn} F^{mn} -  D_m\gs  D^m\gs 
  -\frac{4}{r^2} \gs^2 + \cdots \right ]~,\label{class-vect-act}
  \eea
    we find an $r$-dependent mass-term for $\gs$.  The Lagrangian density of massless  scalar $\phi$ in curved space is 
   given by
    \bea
     L_{scalar} = D_m \phi D^m \phi + \frac{d-2}{4(d-1)} {\cal R}~\phi^2~,
    \eea
 which is invariant under the Weyl transformation of the metric $g_{mn} \rightarrow e^{2\Omega} g_{mn}$
  and the scalar field $\phi \rightarrow e^{\frac{2-d}{2} \Omega} \phi$. Here ${\cal R}$ is the scalar curvature, which for the $d$-sphere is ${\cal R}= \frac{d(d-1)}{r^2}$. Thus restricting to the case of a five dimensional sphere
    we get
     \bea\label{confmass}
     L_{scalar} = D_m \phi D^m \phi + \frac{15}{4r^2}~\phi^2~.
    \eea
  Hence, the vector multiplet  scalar $\gs$ is massive. 

Next we would like to couple the vector multiplet to a  hypermultiplet in representation $\rm{R}$. The massless
 hypermultiplet is conformal so there is a well-defined prescription  to put it on the sphere. 
  The  Lagrangian density on $S^5$ for the hypermultiplet coupled to the vector multiplet is given by the  
   expression
\bea
&&L_{matter}=\epsilon^{IJ} D_m \bar{q}_I  D^m q_J  -\epsilon^{IJ} \bar{q}_I \gs^2  q_J +
\frac{15}{4r^2} \epsilon^{IJ}  \bar{q}_I  q_J
 -2i \bar{\psi} \slashed{D}\psi-2\bar{\psi} \gs \psi \nn \\
 &&\hspace{2cm}-4 \epsilon^{IJ}\bar{\psi} \gl_I q_{J}-iq_I D^{IJ} q_J~,\label{action-matter-1}
\eea
which contains the conformal mass term  in (\ref{confmass}). 
 A more general mass term can be generated through the standard trick of coupling the hypermultiplet to an auxillary vector multiplet and giving an expectation value to the scalar in the multiplet.  This then leads to the mass term \cite{Hosomichi:2012ek}
 \bea\label{massterm}
 L_{mass} =  - M^2 \epsilon^{IJ} \bar{q}_I q_J + \frac{2i}{r} M t^{IJ} \bar{q}_I q_J - 2M \bar{\psi} \psi\,.
\eea
    Since a vector multiplet scalar is  real,  $M$ is assumed to be real.  As discussed in \cite{Hosomichi:2012ek, Kallen:2012va}  the localisation of the model 
 will require the rotation of the scalar $\sigma \rightarrow i \sigma$ and mass $M \rightarrow iM$, otherwise we will fail to get the correct localisation locus. For example, if we rotate $\sigma$ and do not rotate $M$ then we will be unable to argue that the model is localised at $q=0$.
  Thus one can construct on $S^5$ an $\NN=1$ supersymmetric Yang-Mills theory coupled  to a massive hypermultiplet in representation $\rm{R}$. Due to the reality 
  conditions for the hypermultiplet, the representation $\rm{R}$
    and its complex conjugate $\bar {\rm{R}}$ enter 
    the construction symmetrically.  For further details on general  $\NN=1$ theories on $S^5$
    we refer the reader to \cite{Hosomichi:2012ek}. 
  
It is  natural  to ask if one can  construct a theory on $S^5$ with more supersymmetries.  In flat space the $\NN=1$ theory is enhanced to $\NN=2$ if there is a single massless adjoint hypermultiplet.  
 But by itself there is no canonical way to map the $\NN=2$ on flat space to the sphere.  Our best bet is to look
  at the scalar mass terms in (\ref{massterm})  and look for a value for $M$ where there is enlargement 
   of the global symmetry. Combining the relevant terms in (\ref{confmass}) and (\ref{massterm}) we have 
   \bea
    -\frac{4}{r^2} \gs^2 + (\frac{15}{4r^2} - M^2) \epsilon^{IJ} \bar{q}_I q_J + \frac{2i}{r} M t^{IJ} \bar{q}_I q_J ~.\label{sum-terms}
   \eea
  One can see that there is a special point $M= \frac{1}{2r}$ (or $M=-\frac{1}{2r}$) where the terms (\ref{sum-terms}) become \cite{Kim:2012ava}
  \bea
   - \frac{4}{r^2} \gs^2 +  \frac{3}{r^2}   \bar{q}_1 q^1     +  \frac{4}{r^2} \bar{q}_2 q^2\,,                                                                   
  \eea
  thus indicating an enlargement of the symmetry to $SO(1,2) \times SO(2)$. For the point 
  $M=-\frac{1}{2r}$ the roles of $q^1$ and $q^2$ are interchanged.  This is the largest possible symmetry for
   the sphere. The fact that the symmetry enlargement happens for the massive hypermultiplet is 
   because the  vector multiplet scalar itself is massive.

 \section{Matrix model for $\NN=1$ $5D$ Yang-Mills with matter}
 \label{S-general-matrix}

Using the model discussed in the previous section the perturbative partition function was derived in  \cite{Kallen:2012va} for massless hypermultiplets 
 (see also \cite{Kim:2012ava}). Here we discuss some subtle issues in the definition of the matrix model for $\NN=1$ supersymmetric Yang-Mills theory. We follow the conventions in \cite{Kallen:2012va}, spelling  them  out when necessary. From now on we absorb the radius $r$
  into the integration variable $\phi = -i r\gs$.   In complete analogy to  \cite{Pestun:2007rz}, we must integrate over imaginary $\gs$, and hence real $\phi$, in order  to have a well-defined path integral
   (see \cite{Kallen:2012va} for further details).

 Consider the theory with a semi-simple compact gauge group $G$. We have
an $\NN=1$ vector multiplet and an $\NN=1$ massless hypermultiplet in representation $\rm{R}$ (half-hypermultiplets in 
  representations $\rm{R}$ and $\bar {\rm{R}}$). 
 The partition function of this gauge theory on $S^5$ is given by the  following 
  expression 
     \be
Z&=&\int\limits_{\rm Cartan} [d\phi]~e^{-  \frac{ 8 \pi^3 r}{g_{YM}^2}  \text{Tr}(\phi^2)-\frac{\pi k}{3}\text{Tr}(\phi^3)}  Z_{\rm 1-loop}^{\rm vect} (\phi)    Z_{\rm 1-loop}^{\rm hyper} (\phi) + \mathcal{O} (e^{-\frac{16 \pi^3 r}{g_{YM}^2}})~,\label{vh1loop-intro}
\ee
 where the one-loop contributions are given by the following infinite products
\bea
 Z_{\rm 1-loop}^{\rm vect} (\phi) =\prod\limits_\beta\prod\limits_{t \neq 0}\left( t - \langle \beta,i\phi\rangle   \right)^{(1+\frac{3}{2}t+\frac{1}{2}t^2)}~,\label{vect1-loop-beg}
\eea
 and
\begin{equation}\label{main-form-det}
Z_{\rm 1-loop}^{\rm hyper} (\phi) = \prod\limits_\mu \prod\limits_{t}\left( t - \langle i\phi , \mu\rangle +\frac{3}{2} \right)^{-(1+\frac{3}{2}t+\frac{1}{2}t^2)}~.
\end{equation}
 Here $\beta$ are the roots while  $\mu$ are  the weights in  $\rm{R}$. 
 In this expression everything is taken from  \cite{Kallen:2012va} except for the $\text{Tr} \phi^3$ term. Let us 
  explain its origin and its numerical coefficient. We define the gauge theory with matter 
   and   Chern-Simons terms
   \be
    S= \frac{1}{g_{YM}^2} \int\limits_{S^5} \text{Tr} (F \wedge *F) + \cdots + 
    \frac{ik}{24\pi^2} \int\limits_{S^5}\text{Tr}(
    A \wedge dA \wedge dA) + \cdots~,
   \ee
  where we  have only written  terms relevant for the normalisation. The supersymmetrization of the Chern-Simons term was discussed in \cite{Kallen:2012cs}.  Using the normalisations and conventions
   in \cite{Kallen:2012va}, in particular the relation  $\kappa d\kappa d\kappa = - 8 \rm{vol}_g$ where $\kappa$ is 
   the contact form,   
   we get the expression 
    (\ref{vh1loop-intro}).
  
\subsection{Renormalization of the matrix model}
  
 Ignoring for the moment the Lie algebra structure, the building block for a one-loop contribution 
  is given by the following infinite product
  \be
{\cal P}=  x \prod_{t=1}^\infty  \left ( t+ x \right)^{(1+\frac{3}{2}t+\frac{1}{2}t^2)} \left ( t- x \right)^{(1-\frac{3}{2}t+\frac{1}{2}t^2)}~.\label{infin-prod}
  \ee
  As it stands, this infinite product is divergent. The divergent piece is extracted from the following  
 expression
\be
\log {\cal P} = \sum_{t=1}^\infty \left ( 3x - \frac{x^2}{2} \right ) + {\rm convergent~ part}~.
\ee
 Thus the relevant function can be defined by the following
Weierstrass type formula
\be
S_3(x) = 2\pi e^{-\zeta'(-2)} x e^{\frac{x^2}{4} - \frac{3}{2}x} \prod_{t=1}^\infty \Big ( \left ( 1+ \frac{x}{t} \right)^{(1+\frac{3}{2}t+\frac{1}{2}t^2)} \left ( 1- \frac{x}{t} \right)^{(1-\frac{3}{2}t+\frac{1}{2}t^2)}
 e^{\frac{x^2}{2} - 3x} \Big )~,\label{def-triple-sine}
\ee
  where the divergent piece is subtracted from each term. Indeed the above expression is 
   the definition of the triple sine function in terms of an infinite product \cite{kurokawa}. 
    The use of triple sines for 5D partition functions has been pointed out in
   \cite{Lockhart:2012vp, Imamura:2012bm}. 
   
Alternatively, we can introduce a cut-off in order to regularise the divergence. Cutting  the mode expansion of the divergent part off at $n_0 = \pi \Lambda r$,  
  the regularized  1-loop contribution  becomes
\be
\log ( Z_{\rm 1-loop}^{\rm vect} (\phi) Z_{\rm 1-loop}^{\rm hyper} (\phi) ) =  -\frac{\pi\Lambda r}{2}
 \sum\limits_{\beta} (\langle \beta,i\phi\rangle )^2 +   \frac{\pi\Lambda r}{2}
 \sum\limits_{\mu} ( \langle i\phi , \mu\rangle)^2 +   {\rm convergent~ part}\nn \\
 = \pi \Lambda r \left (C_2 ({\rm adj}) - C_2(R) \right ) \text{Tr}(\phi^2 ) + {\rm convergent~ part}~,
\ee
 where  
 $\Tr (T_A T_B) = C_2(R) \delta_{AB}$ and 
  $\sum\limits_{\mu} ( \langle \phi , \mu\rangle)^2 = 2 C_2(R) \Tr (\phi^2)$. 
   We use the conventions   that  $C_2(R) =1/2$ for the fundamental representation of $SU(N)$ .
  The linear 
   piece disappears since the gauge group is  semi-simple. We see that the divergent 
    piece is proportional to $\text{Tr}(\phi^2)$ and thus can be absorbed into a redefinition of the coupling constant by
\be
 \frac{1}{g_{eff}^2} = \frac{1}{g_{YM}^2} - \frac{\Lambda}{ 8\pi^2} \left (C_2 ({\rm adj}) - C_2(R) \right)~.  
 \ee
 The above renormalisation of Yang-Mills coupling agrees 
  with the known flat space results. 
   This renormalisation has been explicitly derived using  supergraph techniques in  \cite{Flacke:2003ac}.
    Alternatively it can be derived by matching 4D and 5D prepotentials on $\mathbb{R}^4 \times S^1$ following  \cite{Nekrasov:1996cz}.
 
  Hence, the convergent part of the infinite product (\ref{infin-prod}) can be replaced by $S_3(x) e^{-\frac{x^2}{4} + \frac{3}{2} x}$
   up to irrelevant ($x$-independent) constants. 
  This additional exponential factor can be absorbed into the $\text{Tr}(\phi^2)$ term by a further shift in the coupling constant,  which is equivalent 
to  redefining the cut-off by a finite shift. Therefore, we  conclude that we can write the matrix model using 
    the effective coupling $g_{eff}$ and the triple sine functions $S_3$.  
 We can also see that in the case when $C_2 ({\rm adj}) = C_2(R)$, the one-loop answer is automatically finite and 
   does not require any regularisation.

\subsection{Massive hypermultiplet}
 
 Assuming that infinite products are regularised (if  necessary)
  we can rewrite our matrix model as follows in terms of triple sine  functions $S_3(x)$
 \bea
 Z=  \int d\phi e^{-\frac{ 8 \pi^3 r}{g_{YM}^2}  \Tr (\phi^2)-\frac{\pi k}{3}\text{Tr}(\phi^3)} {\rm det}_{Ad} \Big ( S_3(i\phi)  \Big )\
   {\rm det}^{-1}_{R} \Big ( S_3 \left (i\phi + \frac{3}{2} \right  ) \Big )~,
 \eea
  where from now on we assume that $g_{YM}$ is the renormalized coupling.  
 The   triple sine function $S_3(x)$ has the following  symmetry properties,
  \bea
   S_3( - x) = S_3(x+3)~,~~~~~~S_3\left (x+\frac{3}{2} \right ) = S_3\left (- x+\frac{3}{2} \right )~.
  \eea
Assuming the standard normalization for the group generators, $[T_A, T_B] = if_{AB}^{~C} T_C$, the weights are switched from   $\mu$ to $-\mu$ when exchanging representation $\rm{R}$ with $\bar {\rm{R}}$. Hence, we have the following property
    for the one-loop contribution of a massless hypermultiplet
  \bea
    {\rm det}_{R} \Big ( S_3 \left (i\phi + \frac{3}{2} \right  )   \Big )=  {\rm det}_{R} \Big ( S_3 \left (-i\phi + \frac{3}{2} \right  ) \Big )
    = {\rm det}_{\bar{R}} \Big ( S_3 \left (i\phi + \frac{3}{2} \right  ) \Big )~.
  \eea
   Therefore, the representations $\rm{R}$ and $\rm{\bar {R}}$ are automatically symmetrized in the determinants, as is required for a hypermultiplet representation.
     
   Masses for  hypermultiplets can be turned on easily 
    by using the auxiliary $U(1)$ vector multiplet   discussed in the previous section. 
We simply   take a  
      $G \times U(1)$ matrix model, but exclude the integratation over the $U(1)$ direction.  Thus the contribution of massive 
      hypermultiplet is given by
        \bea
 Z=  \int d\phi e^{-\frac{ 8 \pi^3 r}{g_{YM}^2} \Tr (\phi^2)- \frac{\pi k}{3} \Tr (\phi^3)} {\rm det}_{Ad} \Big ( S_3(i\phi)  \Big )\
   {\rm det}^{-1}_{R} \Big ( S_3 \left (i\phi +i m + \frac{3}{2} \right  ) \Big )~,
   \eea
where $m$ is related to the hypermultiplet mass parameter $M$ in (\ref{massterm}) by  $m= -irM$. As we have stressed in section \ref{actions-5Dtheory} the rotation of $M$ to imaginary values is required by  localization.    Using the triple sine symmetry properties,  we find the relation
   \bea
    {\rm det}_{R} \Big ( S_3 \left (i\phi + i m + \frac{3}{2} \right  )   \Big ) =   
     {\rm det}_{\bar{R}} \Big ( S_3 \left (i\phi - i m + \frac{3}{2} \right  )   \Big )\,. 
    \eea
     Thus the partition function with a massive hypermultiplet  can be also written as  
     \bea
  \int d\phi &&e^{-\frac{ 8 \pi^3 r}{g_{YM}^2} \Tr (\phi^2)-\frac{\pi k}{3} \Tr(\phi^3)} {\rm det}_{Ad} \Big ( S_3(i\phi)  \Big )\  \nn \\
 &&  \times {\rm det}^{-1/2}_{R} \Big ( S_3 \left (i\phi + i m +  \frac{3}{2} \right ) \Big ) {\rm det}^{-1/2}_{\bar{R}}\Big ( S_3 \left (i\phi - i m +  \frac{3}{2} \right  )  \Big ) ~. \label{full-matrix-model}
   \eea
  
 \subsection{Large volume limit}
 
 Let us now study  the matrix model in the large volume limit for the $S^5$.
 Let us write the matrix model in the following form
 \be
   \int d\phi~  e^{-{\cal F}}~,
 \ee
 where 
 \be
 {\cal F} = \frac{ 8 \pi^3 r}{g_{YM}^2} \Tr (\phi^2)+\frac{\pi k}{3}\Tr(\phi^3) - \sum_\beta \log S_3( \langle i\phi, \beta\rangle ) + \sum\limits_\mu \log S_3 \Big (\langle i\phi, \mu \rangle + im + \frac{3}{2} \Big )~. \label{full-F-prep}
 \ee
  We can restore the $S^5$  radius  dependence by the rescaling
  $\phi \rightarrow r\phi$ and $m \rightarrow rm$. Using 
 the asymptotic expansion\footnote{In this asymptotic expansion the dots stand for the constant term. This expansion is typically derived  for the region $0< \text{Re} z < 3$ using the integral representation for $S_3(z)$. The case $\text{Re} z=0$ is derived separately via the explicit 
  expression of $S_3$ in terms of polylogs.}  for $| \text{Im} z| \rightarrow \infty$ and $0\leq \text{Re} z < 3$ 
  \be
 \log S_3(z) \sim - \text{sgn} ( \text{Im} z) \pi i \left ( \frac{1}{6} z^3 - \frac{3}{4}z^2 + z + ...   \right )
 \ee
 we obtain the following behaviour
 \be
\frac{1}{2\pi r^3}  {\cal F} =  \frac{ 4  \pi^2}{g_{YM}^2} \Tr (\phi^2)+\frac{ k}{6}\Tr(\phi^3) + 
\frac{1}{12} \Big ( \sum_\beta |\langle \phi, \beta\rangle |^3 - \sum\limits_\mu | \langle \phi, \mu \rangle + m|^3 \Big )+ O(r^{-2})~.
 \ee
  Modulo a constant which was absorbed into the definition  of the coupling in  \cite{Seiberg:1996bd,Intriligator:1997pq}, this expression matches the  exact quantum prepotential
   in the flat-space limit  \cite{Intriligator:1997pq}.   The normalisation in front of the quadratic term can be fixed either by a direct one-loop calculation in  flat space  \cite{Flacke:2003ac} or by matching the 5D superpotential with the 4D superpotential as in  \cite{Nekrasov:1996cz}.
    Notice that $M$ in (\ref{massterm}) must be rotated to the imaginary axis in order for $m$ to match  the mass parameter of the flat space   physical theory    \cite{Intriligator:1997pq}.
 
 \subsection{Well-defined matrix model}
 
We next ask under what conditions the matrix model is well-defined, {\it i.e.} the matrix 
  integral converges. We answer this by going to large $\phi$ and finding if  ${\cal F}$ is a convex positive  function in this limit.
  Taking the limit,   we get the following asymptotic expression
    \be
    {\cal F} =  \frac{ 8 \pi^3 r}{g_{YM}^2} \Tr (\phi^2)+\frac{\pi k}{3}\Tr(\phi^3) + 
\frac{\pi}{6} \Big ( \sum_\beta |\langle \phi, \beta\rangle |^3 - \sum\limits_\mu | \langle \phi, \mu \rangle|^3 \Big )&&\nn \\
 -\frac{\pi}{2} m \sum_\mu \text{sgn} (  \langle \phi, \mu \rangle)  (\langle \phi, \mu \rangle)^2
 - \pi \sum_\beta |\langle \phi, \beta\rangle | - \frac{\pi}{2} \left ( m^2 + \frac{1}{4} \right ) \sum_\mu |\langle \phi, \mu \rangle | + \cdots~,
    \ee
    where $\cdots$ stands for  terms suppressed for large   $\phi$.  Analyzing the  convexity of the above function, it is clear that 
     the cubic terms dominate.  Hence, the analysis is identical to the one presented by \cite{Intriligator:1997pq} and so the same conditions apply.  In some special cases  the cubic terms can cancel each other. 
      For example, this happens in the case of single adjoint hypermutiplet 
       \cite{Kallen:2012zn} or in the case of $USp(2N)$ theory with specific matter content
   \cite{Jafferis:2012iv}.     
       
  \subsection{Decoupling of a massive hypermultiplet} 
  
We now consider the behavior of the matrix model as we send the hypermultiplet mass to infinity.  For large $|m|$ the leading terms in (\ref{full-F-prep}) are
  \be\label{F:def}
  {\cal F} = \frac{ 8 \pi^3 r}{g_{YM}^2} \Tr (\phi^2)+\frac{\pi k}{3}\Tr(\phi^3) - \sum_\beta \log S_3( \langle i\phi, \beta\rangle ) - \text{sgn} (m) \frac{\pi}{2} 
   \sum\limits_\mu \left ( \frac{1}{3} ( \langle \phi, \mu \rangle )^3 + m (\langle \phi, \mu \rangle)^2 \right )\,.\nn
 \ee
The two last terms can be absorbed by a redefinition of $k$ and $g_{YM}$.
  
    To see this explicitly, we note that
    \be\label{trace3}
     \text{Tr} (T_A T_B T_C + T_A T_C T_B) = C_3( R ) d_{ABC}~,
    \ee
  where we used the following relation
  \be\label{C3}
  \sum\limits_\mu   ( \langle \phi, \mu \rangle )^3 = C_3 ( R ) \text{Tr} (\phi^3)~.   
  \ee
 The coefficient $C_3$ satisfies $C_3( \bar{R} ) = - C_3( R )$, hence it is zero for real representations.  For the lower complex representations  in $SU(N)$ it is  $1$ for the fundamental,  $N-4$ for the antisymmetric, and $N+4$ for the symmetric representations.  
    Hence, from (\ref{trace3}) and (\ref{C3}) we get 
    \be
     k_{eff} = k  - \text{sgn}(m) \frac{C_3 ( R )}{2}\,,
    \ee
   reproducing the result in  \cite{Seiberg:1996bd,Intriligator:1997pq}.
Similar analysis of the quadratic terms gives the formula 
\be
 \frac{r}{g_{eff}^2} = \frac{r}{g_{YM}^2} - \frac{|m|}{ 8\pi^2} C_{2}(R)~.
\ee

\section{$\NN=1^*$ $5D$ super Yang-Mills}
\label{adjoint-5D-YM}

 From now on we concentrate on   a single adjoint hypermultiplet with mass parameter $m$, which we refer to as $\NN=1^*$  super Yang-Mills.   We also assume that the gauge group is $SU(N)$.
 
 \subsection{The free energy}
 In order to analyse the matrix model (\ref{full-matrix-model}) we  use an alternative representation of the triple sine $S_3(z)$ as defined in (\ref{def-triple-sine}). Namely we can rewrite   $S_3(z)$
 as 
 \bea
  S_3(z) = 2 e^{-\zeta'(-2)} \sin (\pi z) ~e^{\frac{1}{2} f(z)} ~e^{\frac{3}{2} l(z)}~,
 \eea
 where $l(z)$ is the function introduced by Jafferis \cite{Jafferis:2010un}
 \be
l(z)=-z\log\left(1-e^{2\pi i z}\right)+\frac{i}{2}\left(\pi z^2+\frac{1}{\pi}\mathrm{Li}_2(e^{2\pi i z})\right)
-\frac{i \pi}{12}
\label{l:function}
\ee
and $f(z)$ is the function introduced in \cite{Kallen:2012cs}
\be
f(z)=\frac{i\pi z^3}{3}+z^2 \log\left(1-e^{-2\pi i z}\right)+\frac{i z}{\pi}\mathrm{Li}_{2}\left(e^{-2\pi i z}\right)
+\frac{1}{2\pi^2}\mathrm{Li}_{3}\left(e^{-2\pi i z}\right)-\frac{\zeta(3)}{2\pi^2}~.
\label{f:function}
\ee
  Using this representation the matrix model (\ref{full-matrix-model}) can be rewritten as 
   follows
\begin{eqnarray}
\nonumber
Z =\int\limits_{Cartan}\left[d\phi\right]e^{-\frac{ 8 \pi^3 r}{g_{YM}^2}\mathrm{Tr}(\phi^2)}&&\prod_{\beta}
(\sin(\pi\langle\beta, i\phi\rangle) e^{-
\frac{1}{4}l(\frac{1}{2}-i m -\langle\beta, i\phi\rangle)-
\frac{1}{4}l(\frac{1}{2}-i m +\langle\beta, i\phi\rangle)}
\\
&\times & e^{\frac{1}{2}f(\langle\beta, i\phi\rangle)-
\frac{1}{4}f(\frac{1}{2}-i m -\langle\beta, i\phi\rangle)-
\frac{1}{4}f(\frac{1}{2}-i m +\langle\beta, i\phi\rangle)} + \cdots~,
\label{main-matrix123}
\end{eqnarray}
 where $\cdots$ stands for the contribution from instantons.  Our main focus is on Yang-Mils, so we have  dropped the Chern-Simons term.
 Using the t' Hooft coupling constant 
$$\lambda=\frac{g_{YM}^2 N}{r}~,$$
and taking the large  $N$ limit at fixed $\lambda$, all instanton contributions are suppressed and 
 the partition function in (\ref{main-matrix123}) reduces to the matrix integral. Specialising to $SU(N)$
 we rewrite the partition function (\ref{main-matrix123}) in terms of the eigenvalues $\phi_i$
and end up with  the following matrix model
\begin{eqnarray}
\nonumber
& &Z \sim  \int \prod_{i=1}^{N}d\phi_{i}\exp\left(-\frac{ 8 \pi^3 N}{\lambda}\sum\limits_{i}\phi_{i}^2+
\sum\limits_{j\neq i}\sum\limits_{i}\biggl[\log\left[\sinh(\pi(\phi_i-\phi_j))\right]
\biggr.\right.
\\
\nonumber
& & 
-\frac{1}{4}l\left(\frac{1}{2}-i m+i(\phi_i-\phi_j)\right)
 -\frac{1}{4}l\left(\frac{1}{2}-i m-i(\phi_i-\phi_j)\right)+ \frac{1}{2}f(i(\phi_i-\phi_j))-
\\
& &\left.\left.
-\frac{1}{4}f\left(\frac{1}{2}-i m+i(\phi_i-\phi_j)\right)-
\frac{1}{4}f\left(\frac{1}{2}-i m-i(\phi_i-\phi_j)\right)\right]\right)~. 
\label{partition:function}
\end{eqnarray}
  In order to proceed further let us review the relevant properties of the $f$- and $l$-functions. 
These are:
\begin{enumerate}
 \item $l(z)$ is an odd function and $f(z)$ is an even function 
 $$l(z)=-l(-z),\qquad f(z)=f(-z)$$
\item 
  The derivatives of the functions are given by the following simple expressions
    \be
\frac{d f(z)}{dz}=\pi z^2 \cot(\pi z)\, ;\qquad \frac{d l(z)}{dz}=-\pi z \cot(\pi z) \,;
\label{f:l:derivative}
\ee
\item 
  The asymptotic behavior of the functions are given by
   \begin{eqnarray}
 \lim_{|x|\to\infty} \mathrm{Re}f\left(\frac{1}{2}+i x\right) =  -\frac{\pi}{3}|x|^3+\frac{\pi}{4}|x|\,;  &
 \lim_{ x \to \infty } \mathrm{Im}f\left(\frac{1}{2}\pm i x\right)  =  \pm\frac{\pi}{2}x^2\, ;
\nonumber
\\
 \lim_{|x|\to\infty} \mathrm{Re}l\left(\frac{1}{2}+i x\right)=  -\frac{\pi}{2}|x|\,;  &
 \lim_{x\to\infty} \mathrm{Im} l\left(\frac{1}{2}\pm i x\right) = \mp\frac{\pi}{2}x^2\, ;
 \label{asymptotes}
 \\
 \nonumber
  \lim_{|x|\to\infty} \mathrm{Re}f\left(i x\right) =  -\frac{\pi}{3}|x|^3\,;  &
 \mathrm{Im} f\left(i x\right)  = 0\, .
\end{eqnarray} 
\end{enumerate}

In the large  $N$ limit the partition function is dominated by the saddle point.  Using the derivatives in (\ref{f:l:derivative})   the $\phi_i$ satisfy 
\begin{eqnarray}
\nonumber
\frac{ 16 \pi^3 N}{\lambda}\phi_i&=& \pi \sum\limits_{j\neq i}\Bigg[\left(2- (\phi_i-\phi_j)^2\right)\coth(\pi(\phi_i-\phi_j))
\\
&&\qquad\qquad+\frac12\left(\frac{1}{4}+(\phi_i-\phi_j-m)^2\right)\tanh(\pi(\phi_i-\phi_j- m))\nn\\
&&\qquad\qquad+\frac12\left(\frac{1}{4}+(\phi_i-\phi_j+ m)^2\right)\tanh(\pi(\phi_i-\phi_j+ m))\Bigg]\,.
\label{eom}
\end{eqnarray}
For weak coupling where $\la\ll1$, (\ref{eom}) reduces to
\be
\frac{16 \pi^3 N}{\lambda}\phi_i&\approx& 2\sum_{j\ne i}\frac{1}{\phi_i-\phi_j}~,
\ee
whose solution has the usual Wigner distribution
\be\label{rho-weak}
\rho(\phi)\equiv\frac{1}{N}\frac{dn}{d\phi}=\frac{2}{\pi\phi_0^2}\sqrt{\phi_0^2-\phi^2}\qquad \phi_0=\sqrt{\frac{\la}{4 \pi^3}}~,
\ee
where 
\be
\int \rho(\phi)d\phi=1~.
\ee
In this limit the free energy has the typical weak coupling form
\be
F=-\log Z\approx -N^2\log \sqrt{\la}\,.
\ee

In the opposite limit where $\la\gg1$, we can assume that $|\phi_i-\phi_j|\gg1$ if we further assume that $|m| \ll\la$.  In this case the saddle point equation simplifies to
\be
\frac{16 \pi^3 N}{\lambda}\phi_i=\pi
\left(\frac{9}{4}+ m^2\right)\sum\limits_{j\neq i}\mathrm{sign}(\phi_i-\phi_j)\,.
\label{eom:strong}
\ee
Assuming that the eigenvalues $\phi_i$ are ordered, we get the solution
\be
\phi_i=\frac{\left(9+4m^2\right)\lambda}{64\pi^2 N}(2i-N)~,
\label{solution:strong}
\ee
which corresponds to an eigenvalue density
\be\label{rho-strong}
\rho(\phi)&=&\frac{32 \pi^2 }{\left(9+4 m^2\right)\lambda}\quad |\phi|\le\phi_m~,\qquad \phi_m=\frac{\left(9+4m^2\right)\lambda}{64\pi^2}\nn\\
&=&\qquad 0\qquad \qquad |\phi|>\phi_m~.
\ee

Using the asymptotic expressions in (\ref{asymptotes}), the matrix model (\ref{main-matrix123})  at strong coupling can be simplified to
\be
Z\sim\int\prod_{i} d\phi_{i} e^{-\frac{ 8\pi^3N}{\lambda}\sum\limits_{i}\phi_{i}^2+\frac{\pi}{2}
\left(\frac{9}{4}+m^2\right)\sum\limits_{j\neq i}\sum\limits_{i}|\phi_i-\phi_j|}~.
\label{partition:strong}
\ee
Substituting the saddle point solution (\ref{solution:strong}) back into (\ref{partition:strong}), 
we find the free   energy,
\be\label{FE}
 F \equiv -\log Z\approx -\frac{g_{YM}^2 N^3}{ 96\pi r}\left(\frac{9}{4}+m^2\right)^{2}
 \,,
 \label{matrix:free:energy}
\ee
 where we  used the approximations
\be
\sum\limits_{i=1}^{N}(2i-N)^2\approx \frac{1}{3}N^3\,,\qquad
\sum\limits_{j\neq i}\sum\limits_{i=1}^{N}|i-j|\approx \frac{1}{3}N^3\,.
\ee
Hence, we see that going from weak to strong coupling the free energy crosses over from $N^2$ to $N^3$ behavior.  At the point $m=\frac{1}{2}$
 the free energy 
 is 
 \be\label{FE11}
 F  = -\frac{ 25g_{YM}^2 N^3}{ 384\pi r} ~.
 \label{matrix:free:energy222}
\ee

It is interesting to see how the system evolves as $m$ approaches a large value.  In this limit we expect the system to reduce to $\NN=1$ SYM with no hypermultiplets.  Indeed, if we assume that $m\gg|\phi_i-\phi_j|$ for all $i$ and $j$, the saddle point equation reduces to
\begin{eqnarray}\label{Largemass}
\frac{16 \pi^3 N}{\lambda}\phi_i&=& \pi \sum\limits_{j\neq i}\Bigg[\left(2- (\phi_i-\phi_j)^2\right)\coth(\pi(\phi_i-\phi_j))+2m (\phi_i-\phi_j)\Bigg]\nn\\
&=& 2\pi m N\phi_i+\pi \sum\limits_{j\neq i}\left(2- (\phi_i-\phi_j)^2\right)\coth(\pi(\phi_i-\phi_j))\,,
\ee
where we used that $\sum_i\phi_i=0$.  We can then reexpress (\ref{Largemass}) as
\be
\frac{16 \pi^3 N}{\lambda_{eff}}\phi_i&=& \pi \sum\limits_{j\neq i}\left(2- (\phi_i-\phi_j)^2\right)\coth(\pi(\phi_i-\phi_j))\,,
\ee
which is the saddle point equation for the $\NN=1$ system with no hypermultiplets and an effective 't Hooft coupling
\be
\frac{1}{\la_{eff}}=\frac{1}{\lambda}-\frac{m}{8\pi^2}\,.
\ee
Note that this is consistent with our consideration in section \ref{S-general-matrix}   and the
 explicit one-loop flat result where $m$ is regarded as UV-regulator.

\subsection{Supersymmetric Wilson loops}
\label{Wilson-loop}

Supersymmetric Wilson loops can be evaluated using localization since they preserve some of the supersymmetries.   In 5D flat space such loops were first considered in \cite{Young:2011aa}.  Supersymmetric Wilson loops  were also considered in \cite{Kim:2012qf,Assel:2012nf}. 

On $S^5$, the supersymmetric loop must  go along an 
 $S^1$ fiber when the  $S^5$ is viewed as an $S^1$-fibration over $\mathbb{C}P^2$.  Thus proceeding in analogy with
 \cite{Pestun:2007rz} the expectation value of supersymmetric Wilson loop is given by 
  the following matrix model expectation value
\be
\langle W\rangle=\frac{1}{N}\langle \Tr e^{2\pi\phi_{i}}\rangle~.
\ee
Taking into account the consideration from previous subsections we first observe that 
in the large $N$ limit the $\sum\limits_{i}e^{2\pi\phi_{i}}$ term has a negligible back-reaction on the position of the 
saddle point in  (\ref{solution:strong}).  Thus at strong coupling $\lambda$ we have to evaluate 
 the integral
 \begin{eqnarray}
\langle W\rangle\sim\frac{1}{N}\int\prod_{i} d\phi_{i}\sum\limits_{i}e^{2\pi\phi_{i}} e^{-\frac{8 \pi^3N}{\lambda}\sum\limits_{i}\phi_{i}^2+\frac{\pi}{2}
\left(\frac{9}{4}+m^2\right)\sum\limits_{j\neq i}\sum\limits_{i}|\phi_i-\phi_j|}~.
\end{eqnarray}
Thus at large $N$ the  Wilson loop expectation value is well-approximated by the following
integral
\be
\langle W\rangle = \int d\phi \rho(\phi)e^{2\pi\phi}
\ee
where $\rho(\phi)$ is the density of eigenvalues. 

For weak coupling we can approximate the integral as
\be
\langle W\rangle \approx \int d\phi \rho(\phi)(1+2\pi^2\phi^2)=1+\frac{\la}{ 8 \pi}\approx \exp\left({\frac{\la}{8\pi}}\right).
\ee
At strong coupling, using the eigenvalue distribution in (\ref{rho-strong}), we find
\be
\langle W\rangle \approx  \frac{ 32 \pi^2 }{\left(9+4m^2\right)\lambda}\int\limits_{-\phi_m}^{\phi_m}e^{2\pi\phi}d\phi\sim
\exp\left({\frac{\lambda}{ 8\pi}\left(\frac{9}{4}+m^2\right)}\right),	
\label{wilson:loop:matrix}
\ee
where we have omitted the prefactor, as only the exponential term is important for us.  Comparing the weak and strong coupling results, we see that at large $\lambda$ the effect on the Wilson loop is to rescale $\lambda$ by the factor  $\left(\frac{9}{4}+m^2\right)$. 

\subsection{Quivers}

The above results can be straightforwardly generalized to a $\mathbb{Z}_k$ quiver of the $\NN=1^*$ theory \cite{Kallen:2012zn}.  Here we have an $SU(N/k)^k$ gauge group and  equal mass hypermultiplets  in the bifundamental representations,
$(N/k,\overline{N/k},1,\dots 1)$, $(1,N/k,\overline{N/k},1,\dots )$, {\it etc.}.   The eigenvalues of (\ref{eom}) are split into $k$ groups $\psi^{(r)}_i$, where $r=1,\dots,k$, $i=1,\dots N/k$, and the resulting equations of motion are
\begin{eqnarray}
\frac{ 16\pi^3 N}{\lambda}\psi^{(r)}_i&=& \pi \Big[\sum\limits_{j\neq i}\left(2- (\psi^{(r)}_i-\psi^{(r)}_j)^2\right)\coth(\pi(\psi^{(r)}_i-\psi^{(r)}_j))\nn
\\
&&\qquad+\Bigg(\sum_j\Big[\sfrac14\left(\sfrac{1}{4}+(\psi^{(r)}_i\ms\psi^{(r\ps1)}_j\ms m)^2\right)\tanh(\pi(\psi^{(r)}_i\ms\psi^{(r\ps1)}_j\ms m))\nonumber\\
&&\qquad\qquad+\sfrac14\left(\sfrac{1}{4}+(\psi^{(r)}_i\ms\psi^{(r\ms1)}_j\ms m)^2\right)\tanh(\pi(\psi^{(r)}_i\ms\psi^{(r\ms1)}_j\ms m))\Big]\Big]\nn\\
&&\qquad\qquad\qquad +(m\to -m)\Bigg)\,.
\label{eom-quiver}
\end{eqnarray}
These have a solution where $\psi^{(r)}_i=\psi^{(s)}_i$ or all $r$ and $s$, hence taking the same limits as before we find that the eigenvalues sit at
\be
\psi^{(r)}_i=\frac{\left(9+4m^2\right)\lambda}{ 64 \pi^2 N}(2i-N/k)\,.
\label{solution:strongq}
\ee
Thus, the free energy is
\be\label{FEq}
 F \approx -k\,\frac{g_{YM}^2 N^3}{ 96 \pi rk^3}\left(\frac{9}{4}+m^2\right)^{2}=-\frac{g_{YM}^2 N^3}{96 \pi rk^2}\left(\frac{9}{4}+m^2\right)^{2}
 \,,
\ee

For supersymmetric Wilson loops, the weak coupling behavior parallels the $\NN=1^*$ case with $\lambda$ replaced by $\lambda/k$, hence
\be\label{Wilsonweakq}
\langle W\rangle\approx \exp\left({\frac{\la}{ 8 \pi\,k}}\right)\,.
\ee
At   strong coupling the leading behavior of the Wilson loop is  determined by the top eigenvalue, hence we have
\be
\langle W\rangle \sim
\exp\left({\frac{\lambda}{ 8\pi\,k}\left(\frac{9}{4}+m^2\right)}\right),	
\label{wilson:loop:matrix:q}
\ee
Therefore, we find the same large $\lambda$ rescaling of  the  Wilson loop  as in the $\NN=1^*$ case.

\section{Numerical study of $\NN=1$ $5D$ Yang-Mills}
\label{numerics}

In this section we give  numerical evidence for the analytical approximations used in the previous 
 section. 

 We have been unable to find an exact solution for the saddle-point equation in (\ref{eom}). 
 However, we can look for numerical solutions using an   
idea similar to the one used in \cite{Herzog:2010hf}. Instead of solving a
system of $N$ algebraic equations (\ref{eom}) of the form $-\frac{\partial \cal F}{\partial \phi_i}=0$, where $\cal F$ is 
defined in (\ref{F:def}), we introduce a time dependence for the matrix model eigenvalues $\phi_i (t)$ and solve the ``heat" 
 equation
\begin{equation}
 \tau \frac{d\phi_i}{dt}=-\frac{\partial \cal F}{\partial \phi_i}\,.
 \label{heat}
\end{equation}
At  large time-scales where $t\to\infty$, with an appropriate choice of $\tau$ the solution of (\ref{heat}) 
relaxes and approaches  the solution
of the saddle point equations  $-\frac{\partial \cal F}{\partial \phi_i}=0$.

 Following this approach  
 we eventually reach  the density of
 eigenvalues $\rho(\phi)=\frac{dx}{d\phi}$  shown in Fig.~\ref{density:pic}.  
 Here we show the  densities for two different values of $m$. For comparison, we have superimposed these over the 
 corresponding analytical strong coupling result from (\ref{rho-strong}).  As one can see the analytical and 
numerical solutions coincide  at large $N$,  except  for a small region near the boundaries 
of the distribution.
 \begin{figure}
\begin{center}
  \includegraphics[width=53mm,angle=0,scale=1.3]{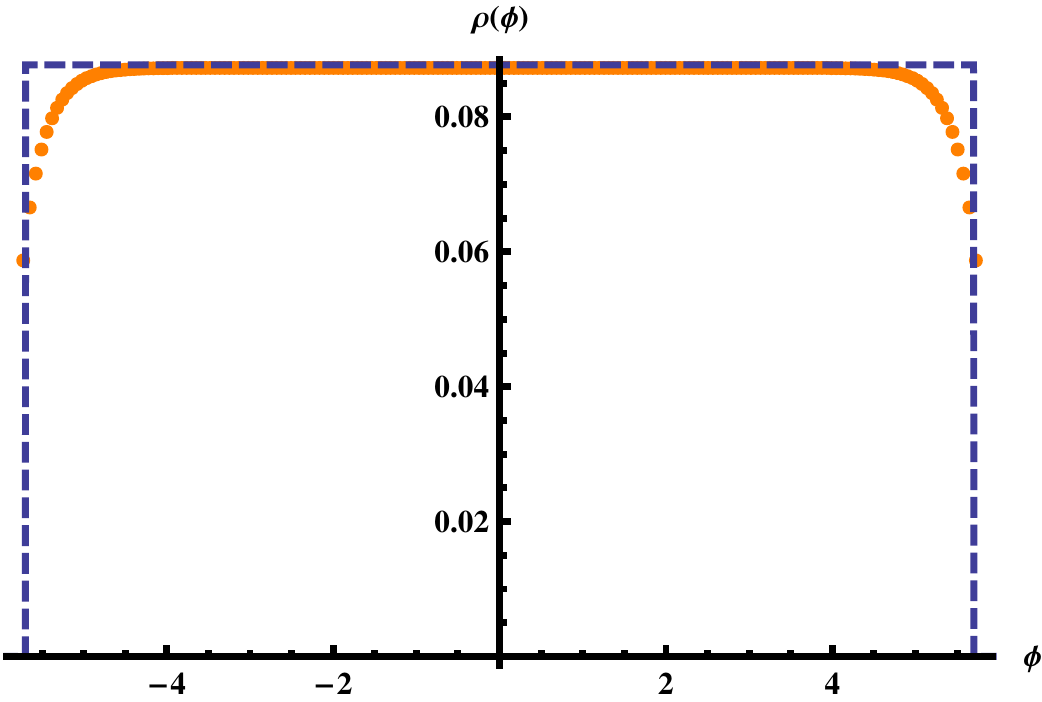}\hspace{5mm}
  \includegraphics[width=50mm,angle=0,scale=1.37]{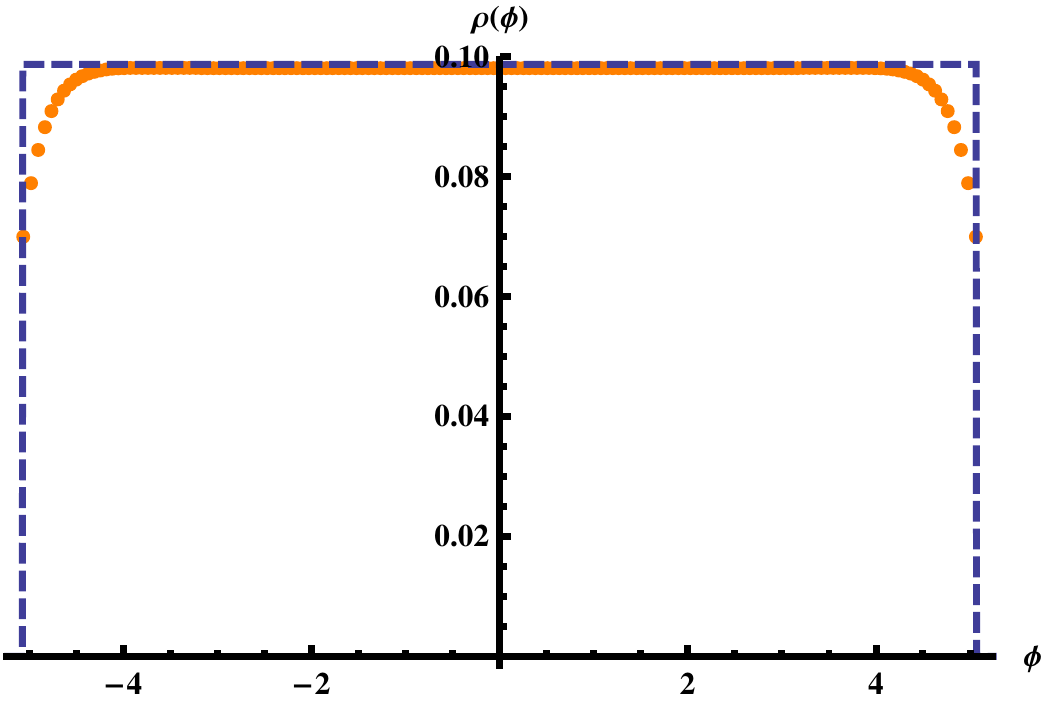}
\end{center}
\caption{Density of eigenvalues $\rho(\phi)$ for $m=0$, $N=200$, 
$\beta\equiv\frac{g_{YM}^2}{r}=2$ \textit{(left)} and $m=\frac{1}{2}$, $N=160$, 
$\beta\equiv\frac{g_{YM}^2}{r}=2$ \textit{(right)}. The dashed blue lines are the analytical strong coupling solutions.}
\label{density:pic}
\end{figure}
In the case of pure $\NN=1$ SYM theory with no hypermultiplets the density of states develops a peak at each end as well as a 
shallow minimum in the center.  This is shown in   Fig.~\ref{density:N=1}. 
  \begin{figure} 
  \begin{center}
    \includegraphics[width=0.48\textwidth]{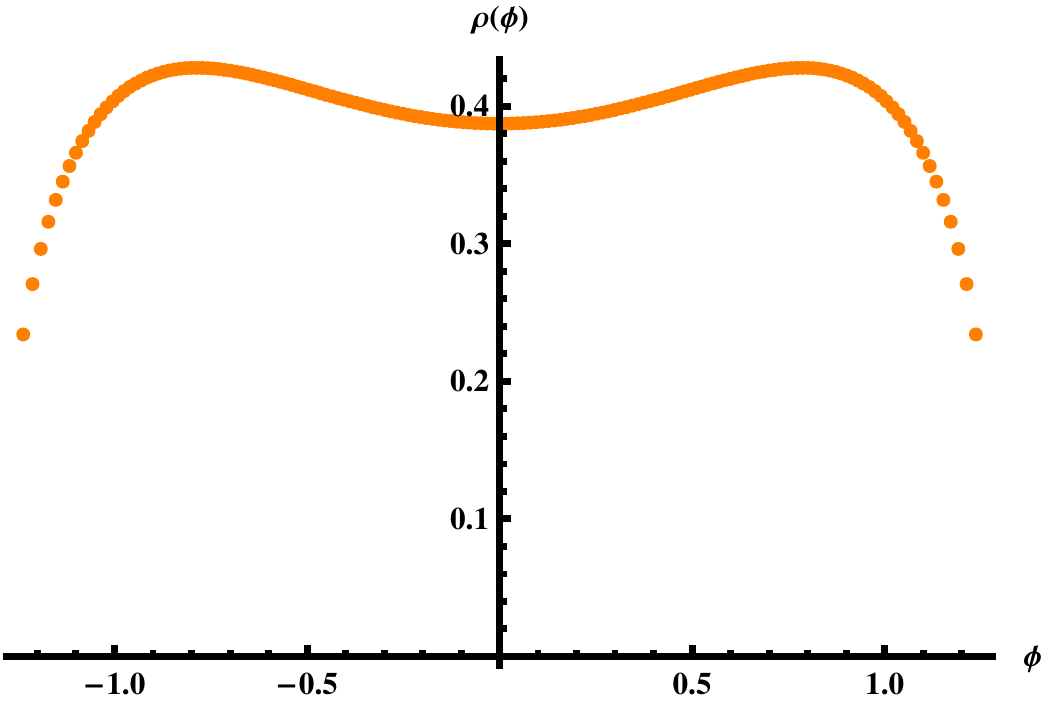}
  \caption{Density of states for pure $\mathcal{N}=1$ SYM   with $\beta=2$. }
  \label{density:N=1}
    \end{center}
\end{figure}

Using these distributions of eigenvalues we   can find the free energy 
 for the matrix model. Results for the energies with different values of 
$m$ and for $\mathcal{N}=1$ are shown on Fig.\ref{free:pic}. The dashed lines 
correspond to the respective  strong coupling results from (\ref{FE}).
\begin{figure}
\begin{center}
  \includegraphics[width=53mm,angle=0,scale=1.3]{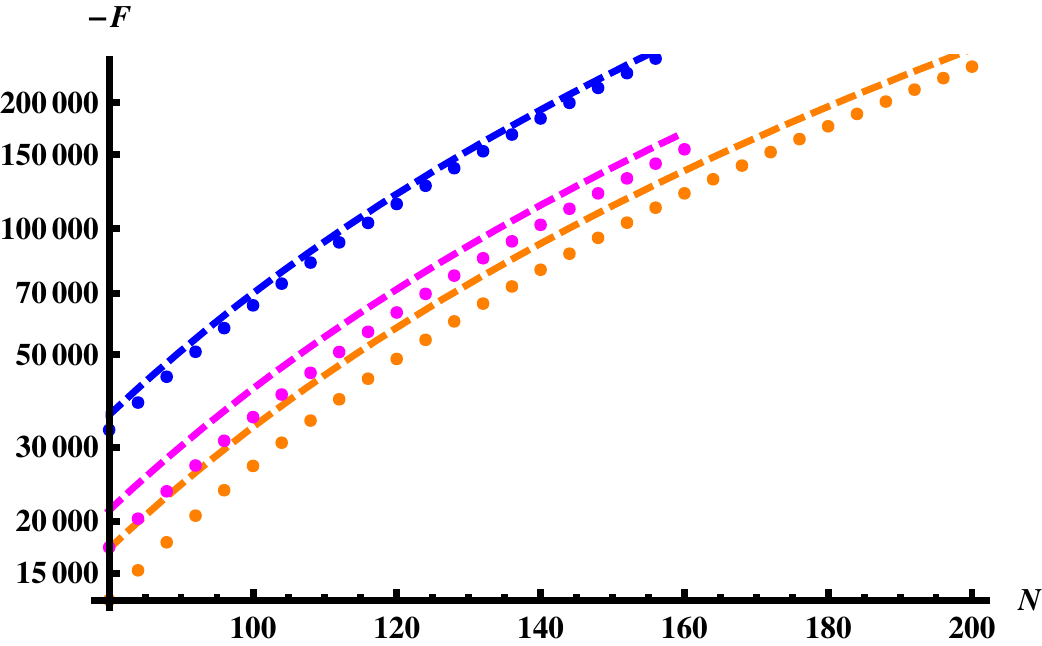}\hspace{5mm}
  \includegraphics[width=50mm,angle=0,scale=1.37]{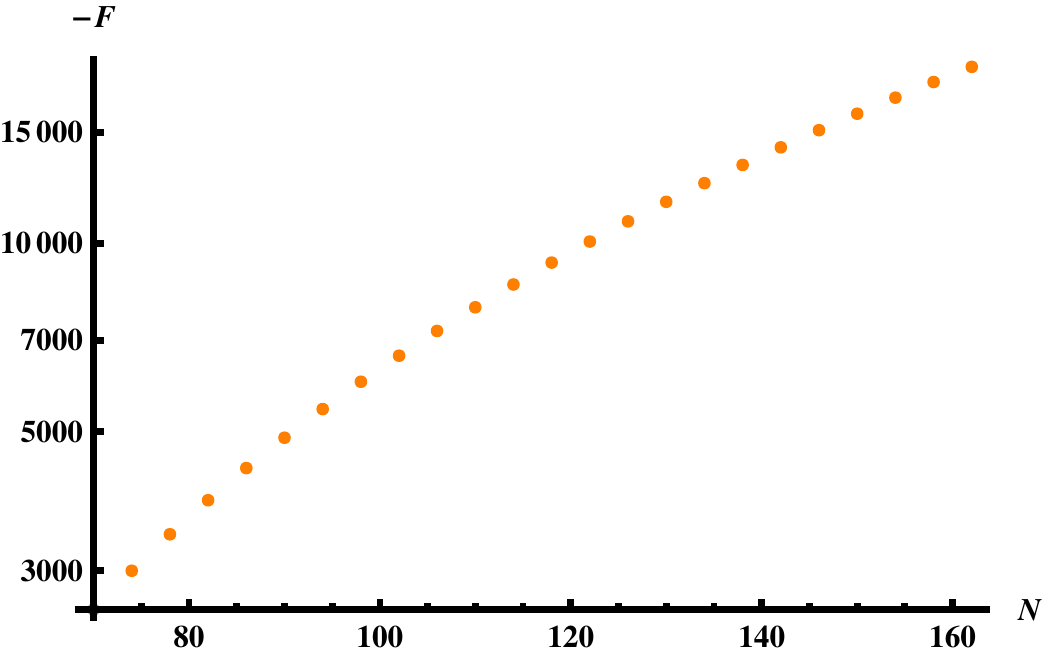}
\end{center}
\caption{\textit{(left)}$N$-dependence of free energies for different values
of hypermultiplet mass $m$: $m=0$ \textit{orange}, $m=\frac{1}{2}$ \textit{purple},
$m=1$ \textit{blue};\hspace{4mm}  \textit{(right)} Free energy $N$-dependence for pure $\NN=1$ SYM.  ($\beta=2$ for all plots.) }
\label{free:pic}
\end{figure}
As the graphs show, the difference between the analytic and numerical results grows as $m$ decreases but converges
as $N$, and hence $\lambda$ is increased. 
 The discrepancy we observe is due to the effect of 
subleading terms in $N$ and $\lambda$.

Having the data for the different $m$ we can fit the  $N$-dependencies 
with a simple polynomial function $F=a_4 N^4+a_3 N^3 + a_2 N^2 + a_1 N + a_0 + ...$. This fit eventually gives the leading $N^3$ term
for general $m$.  The coefficient $a_3$ can  be fit to a polynomial function of $m$,
\begin{equation}
 a_3= c_1 + c_2 m + c_3 m^2 + c_4 m^3 +...
\end{equation}
 Eventually we end up with coefficients 
that are very close to the strong coupling  result in (\ref{FE}).
In  Fig.~\ref{m:depend} we plot the $m$-dependence of the $N^3$ coefficient of the free
 energy. The dashed line  shows the function in    (\ref{FE}).
 The plot shows precise agreement between the  numerical and analytical results.
 
  \begin{figure} 
  \begin{center}
    \includegraphics[width=0.48\textwidth]{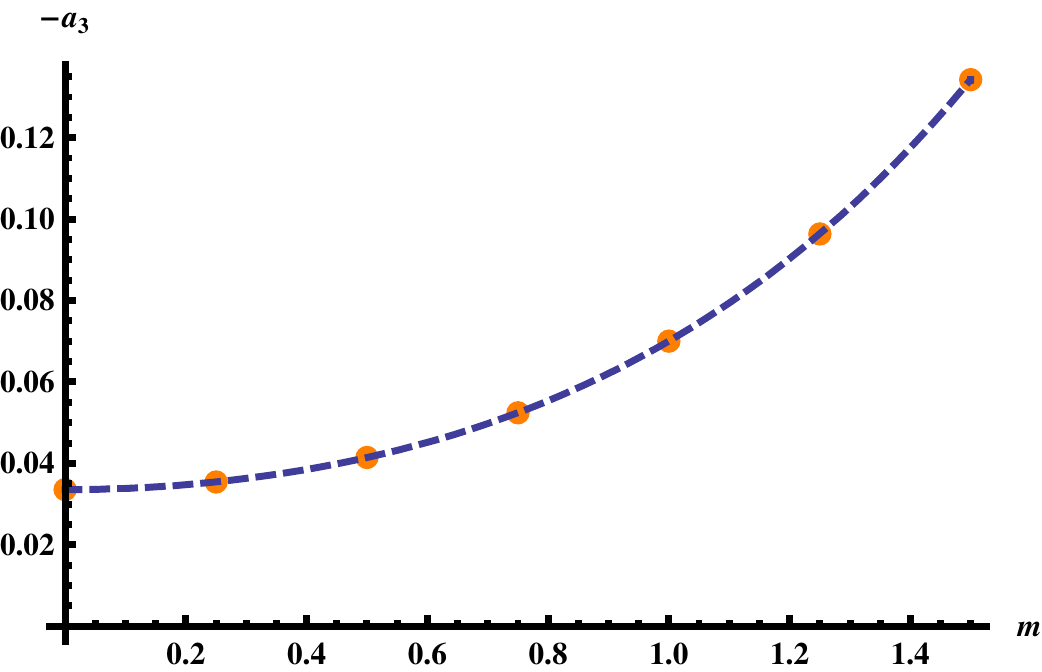}
  \caption{$m$-dependence of $N^3$ coefficient in the matrix model free energy.}
  \label{m:depend}
    \end{center}
\end{figure}

 For pure $\NN=1$ SYM the fit gives the leading term behaving as $N^2$ instead of $N^3$ as in the $\NN=1^*$ theories.

 \section{Supergravity comparisons}
\label{s-sugra}

In this section we compare our   strong coupling results for 5D SYM with the analogous computations in supergravity.  To start we review the supergravity computation of the free   energy given in \cite{Kallen:2012zn}.  We consider supergravity  on $AdS_7\times S^4$ where the $AdS_7$ boundary is $S^1\times S^5$.   The radii of $AdS_7$ and $S^4$ are $\ell$ and $\ell/2$ respectively, where $\ell=2\ell_{pl}(\pi N)^{1/3}$.  We write the $AdS_7$ metric as
\be
ds^2=\ell^2(\cosh^2\rho\, d\tau^2+d\rho^2+\sinh^2\rho\, d\Omega_5^2)~,
\ee
where $d\Omega_5^2$  is the unit 5-sphere metric.  The Euclidean time direction is compactified and has the identification $\tau\equiv \tau+2\pi R_6/r$, while $R_6$ and $r$ are the boundary radii of $S^1$ and $S^5$.

Under the AdS/CFT correspondence, the supergravity classical action equals the free   energy of the boundary field theory.  The action needs to be regulated by adding counterterms \cite{Balasubramanian:1999re,Emparan:1999pm,deHaro:2000xn,Awad:2000aj}.  There can be scheme dependence in the regulation \cite{deHaro:2000xn}, but we will follow the minimal subtraction prescription, which is the normal procedure  when regulating the action.  The full action then has the form
\be
I_{AdS}=I_{\rm{bulk}}+I_{\rm{surface}}+I_{\rm{ct}}\,,
\ee
where
\be
I_{\rm{bulk}}=-\frac{1}{16\pi G_N}\mbox{Vol}(S^4)\int d^7x \sqrt{g}\left(R-2\Lambda\right)\
\ee
is the action in the bulk, $I_{\rm{surf}}$ is the surface contribution and $I_{\rm{ct}}$ contains counterterms written only in terms of the boundary metric and which cancel off divergences in $I_{\rm{bulk}}$.  Newton's constant is given by $G_N=16\pi^7\ell_{pl}^9$ \cite{Maldacena:1997re} . The equations of motion lead to
\be
R-2\Lambda=-\frac{12}{\ell^2}\,,
\ee
which when substituted back into the action gives
 \be\label{bulk}
I_{\rm{bulk}}=-\frac{1}{256\pi^8 \ell_{pl}^9}\left(\frac{\pi^2\ell^4}{6}\right)\frac{2\pi R_6}{r}\pi^3 (-12\ell^5)\int_0^{\rho_0}\cosh\rho\sinh^5\rho\, d\rho=\frac{4\pi R_6}{3\,r} N^3\sinh^6\rho_0\,.\nn\\
\ee
In the limit that $\rho_0\to\infty$ the integral is divergent and corresponds to a UV divergence for the boundary theory.  In terms of an $\epsilon$ expansion of the boundary theory, we make the identification $\epsilon=e^{-\rho_0}$, which then gives
\be
\sinh^6\rho_0=\frac{1}{64}\epsilon^{-6}-\frac{3}{32}\epsilon^{-4}+\frac{15}{64}\epsilon^{-2}-\frac{5}{16}+{\rm O}(\epsilon^2)\,.
\ee
The surface term contributes to the divergent pieces, but not the finite part of (\ref{bulk}), while
the effect of the counterterm with minimal subtraction is to cancel off the divergent pieces.  Hence, we find \cite{Emparan:1999pm}
\be\label{IAdS}
I_{AdS}=-\frac{5\pi R_6}{12\,r}N^3~.\label{sugra-final}
\ee

We next consider the supergravity calculation \cite{Berenstein:1998ij} for the Wilson loop which is related to the extremized world-volume of the membrane
\be
\langle W\rangle\sim e^{-T^{(2)}\int dV}\,,
\ee
where $T^{(2)}=\frac{1}{(2\pi)^{2}l_{p}^{3}}$ is the tension of the M2 brane.  The M2 brane is chosen to  wrap the Euclidean time direction and the equator of $S^5$.  The third direction falls in from the boundary into the bulk.  
Hence, the M2 brane volume is given by
\be
\int dV=l^{3}\int\limits_{0}^{\frac{2\pi R_{6}}{r}} d\tau
\int\limits_{0}^{2\pi}d\phi \int\limits_{0}^{\rho_{0}}d\rho\sinh(\rho)\cosh(\rho)
\ee
Using the same UV cutoff as in (\ref{bulk}) we find
\be
T^{(2)}\int dV=\frac{\pi N R_{6}}{r}\left(\frac{1}{\epsilon}-2+\epsilon\right)\,.
\ee
As in the case of the action, the integral needs to be regulated.  Using minimal subtraction again, the result for the regulated Wilson loop is
\be
\langle W\rangle\sim \exp\left(\frac{2\pi N R_{6}}{r}\right)
\label{Wilson:loop:sugra}
\ee

 Comparing this to the strongly coupled result in (\ref{wilson:loop:matrix}), we find that we should set
\be\label{WLagree}
R_6=\frac{g_{YM}^2}{16\pi^2}\left(\frac{9}{4}+m^2\right)\,.
\ee
in order for them to agree.  
  At the $m=1/2$ point the relation is
 \bea
 R_6=\frac{g_{YM}^2}{16\pi^2} \frac{5}{2}~.
 \eea
 Using this relation we see that  (\ref{IAdS}) agrees with (\ref{matrix:free:energy222}).
 
We can also compare supergravity results for the quiver.  The effect of the quiver on the supergravity computation is to replace the $S^4$ with $S^4/\mathbb{Z}_k$.  This reduces the volume factor of the $S^4$ by a factor of $k$, hence
\be\label{IAdSk}
I_{AdS}=-\frac{5\pi R_6}{12\,r\,k}N^3~.\label{sugra-final}
\ee
Comparing the supergravity Wilson loop in (\ref{Wilson:loop:sugra}) to the quiver Wilson loop in (\ref{wilson:loop:matrix:q}), we see that
\be\label{WLagreeq}
R_6=\frac{g_{YM}^2}{ 16\pi^2k}\left(\frac{9}{4}+m^2\right)\,.
\ee
Again choosing $m=1/2$ we have agreement with the free energy in (\ref{FEq}).

\section{Discussion}
\label{summary}

In this paper we have studied the matrix model which corresponds to the full 
perturbative partition function for $\NN=1$ Yang-Mills theory with matter on the five-sphere.
We have  shown how the localized matrix model relates to the flat-space results on the Coulomb branch in \cite{Seiberg:1996bd,Intriligator:1997pq}, 
 \cite{Nekrasov:1996cz}  and explicit one-loop calculations in flat space \cite{Flacke:2003ac}.
This indicates that the prescription we follow, which is analogous to the 4D prescription in \cite{Pestun:2007rz}, is consistent. 
Moreover  we have shown that one can have agreement between the free   energy of the $\NN=1^*$ theory with a specific value of the hypermultiplet mass and the  regularized supergravity action if one incorporates a rescaling of the physical coupling  as well as a Euclidean rotation of the mass-parameter.  The Euclidean rotation of the mass-parameter is needed  to localize the partition function. 

 An important open problem is to better understand the Euclidean version of the 6D $(2,0)$ theory, or if it does not exist, an appropriate substitute.  This includes its symmetries and 
its  reduction to the Euclidean 5D model, which is required for localization. This may provide a stronger argument for the calculations presented in this paper.  
  
\vfill\eject
\bigskip\bigskip

\noindent{\bf\Large Acknowledgement}:
\bigskip

\noindent  We thank  Seok Kim,  Vasily Pestun,  Jian Qiu and  Konstantin Zarembo  for useful discussions
 on this and related subjects.    This  research is supported in part by
Vetenskapsr\aa det under grants \#2011-5079 and \#2012-3269.  J.A.M thanks the
CTP at MIT   for kind
hospitality  during the course of this work.
\\
\\
\bibliographystyle{utphys}
 
 \providecommand{\href}[2]{#2}\begingroup\raggedright\endgroup

\end{document}